\documentclass[12pt]{article}
\usepackage{times}
\usepackage{geometry}
\usepackage[utf8]{inputenc}
\usepackage{enumitem,amssymb}
\usepackage{ragged2e}
\usepackage{setspace}
\usepackage{graphicx}
\newlist{thematic}{itemize}{8}
\setlist[thematic]{label=$\square$}
\usepackage{pifont}
\newcommand{\cmark}{\ding{51}}%
\newcommand{\done}{\rlap{$\square$}{\raisebox{2pt}{\large\hspace{1pt}\cmark}}%
\hspace{-2.5pt}}

\newcommand\apj{The Astrophysical Journal}%    % Astrophysical Journal ++
\newcommand\nat{Nature }%  % Nature 
\begin{document}
\pagenumbering{gobble}
\RaggedRight
\noindent {\fontsize{16}{20} \selectfont Astro2020 Science White Paper}
\begin{center}
{\fontsize{24}{32}\selectfont Radio Observational Constraints on}
\vspace{0.1cm}
{\fontsize{24}{32}\selectfont Turbulent Astrophysical Plasmas}
\end{center}

\vspace{0.3cm}

\normalsize

%\RaggedRight
\noindent \textbf{Thematic Areas:} \hspace*{60pt} $\square$ Planetary Systems \hspace*{10pt} $\square$ Star and Planet Formation \hspace*{20pt}\linebreak
$\square$ Formation and Evolution of Compact Objects \hspace*{31pt} $\square$ Cosmology and Fundamental Physics \linebreak
  \done Stars and Stellar Evolution \hspace*{1pt} $\square$ Resolved Stellar Populations and their Environments \hspace*{40pt} \linebreak
  $\square$    Galaxy Evolution   \hspace*{45pt} $\square$             Multi-Messenger Astronomy and Astrophysics \hspace*{65pt} \linebreak

\justifying
  
\noindent \textbf{Principal Author:} \\
Tim Bastian, National Radio Astronomy Observatory \\
Email: tbastian@nrao.edu.edu \\
Phone: (434) 296-0348 \\

\noindent \textbf{Co-authors:} \\
James Cordes, Cornell University \\
Justin Kasper, University of Michigan \\
Adam Kobelski, West Virginia University \\
Kelly Korreck, Harvard Smithsonian Center for Astrophysics \\
Gregory Howe, University of Iowa\\
Steven Spangler, University of Iowa\\
Chadi Salem, University of California at Berkeley/Space Science Lab\\
Angelos Vourlidas, Johns Hopkins University, Applied Physics Lab\\

\noindent {\bf Abstract} Remarkable progress has been made in understanding turbulent astrophysical plasmas in past decades including, notably, the solar wind and the interstellar medium. In the case of the solar wind, much of this progress has relied on in situ measurements from space-borne instruments. However, ground-based radio observations also have played a significant role and have the potential to play an even bigger role. In particular, using distant background sources (quasars, pulsars, satellite beacons) to transilluminate the foreground corona and solar wind, a variety of radio propagation phenomena can be used to map plasma properties of the solar corona and heliosphere, as well as the warm interstellar medium.  These include angular broadening, interplanetary and interstellar scintillations, and differential Faraday rotation. These observations are highly complementary to in situ observations of the solar wind, and could be a mainstay of investigations into turbulence of the ISM. We point out that the {\sl Next Generation Very Large Array} (ngVLA) fulfills all the requirements necessary to exploit radio observations of astrophysical turbulence fully. 

\newpage

\noindent\textbf{\large Introduction}

Most scientists interested in the field would recognize four major topics in plasma astrophysics. All of these topics concern processes in collisionless plasmas, characterized by collisional mean free paths far greater than important length scales in the plasma. They are (1) the nature of plasma turbulence, including the mechanisms for generation, cascade and dissipation, (2) the nature of magnetic reconnection, including the distinct roles played by electron and ion dynamics, (3) the structure of shock waves, including an understanding of how surfaces of discontinuity can arise on scales much smaller than the collisional scale, and (4) the mechanisms responsible for charged particle acceleration. In this white paper we focus on topic (1) in the context of the solar wind and the interstellar medium, for which radio observations can provide unique insights into the nature of the astrophysical plasmas and their turbulent properties. Separate white papers by {\bf Chen et al.}, {\bf Bastian et al.}, and by {\bf Gary et al.} discuss aspects of topics 2, 3, and 4, respectively. 

Plasma turbulence in the astrophysical context has largely focused on understanding turbulence in the solar wind and in the interstellar medium. Remarkable progress has been made in understanding the nature of the solar wind over the past fifty years (see, e.g., [5]) but mysteries remain. It is widely believed that the solar wind undergoes spatially extended heating, which plays a significant role in solar wind acceleration. Several heating mechanisms have been considered in detail including magnetic reconnection, wave dissipation, and turbulence. As such, the heating and acceleration of the solar wind falls within the domain of the major topics in plasma astrophysics listed above. A detailed understanding of these processes nevertheless remains elusive and fundamental questions remain germane: 

\begin{itemize}
\item how are is the corona  and solar wind heated and how is the solar wind accelerated? 
\item how does strong MHD turbulence develop and evolve in the solar wind? 
\item what are the turbulence dissipation mechanisms? 
\item what role does solar turbulence play in mediating energy deposition and transport? 
\item how does solar wind turbulence modulate the transport of energetic particles (both solar energetic particles and cosmic rays)? 
\end{itemize}

These questions are relevant to other stars, of course. More broadly, the loss of mass and angular momentum due to outflows affect the evolution of a given star [20], impact exoplanetary environments (e.g., [8]), can profoundly influence the interstellar medium in its neighborhood [30], In order to address questions posed by the solar wind and their implications for other stars, our understanding of the solar wind must be placed on a comprehensive observational footing. With the launch of {\sl Parker Solar Probe} and soon, {\sl Solar Orbiter}, the inner heliosphere is a new frontier. For the first time, measurements will be made of the solar wind and solar wind turbulence close to the source. Future missions such as the concept for {\sl Helioswarm} will open an additional avenue of inquiry -- into the multi-scale nature of plasma turbulence. By their nature all space-based missions sample discrete locations. Many locations - the high latitude solar wind or the inner corona, for example - are not accessible to {in situ} measurements. A broader context will be needed within which to interpret space-based observations. 

Most of the interstellar medium (ISM) is partially or fully ionized. The largest component by volume, the warm ionized medium (WIM) is essentially completely ionized. It is also arguably the best-diagnosed plasma outside of the solar system, and it must share important physical processes with the solar wind and solar corona. A review of the properties of the WIM, as determined by a variety of optical and radio observations is given in [11]. The WIM also possesses plasma turbulence, which is primarily diagnosed by the effect of its density fluctuations on radio wave propagation. A review of the techniques and results of these measurements is given by 27, and one of the most important results, the evidence for a large inertial subrange spectrum of turbulence in the ISM, is presented in [1]. Remarkably, recent {\sl in situ} measurements of ISM turbulence by Voyager 1 [21] show good agreement with earlier work. Some of the main issues in the study of the plasma in the WIM are the mechanisms for generation of the turbulence, the magnitude of turbulent dissipative heating, turbulent contributions to transport coefficients such as thermal conductivity and resistivity, and the effect on the propagation and acceleration of cosmic rays.

Radio propagation phenomena have been studied for many years as probes of both the solar wind and the ISM. These studies have demonstrated their utility in constraining key properties of these media. However, a barrier to exploiting them more effectively has been the lack of a ground-based radio interferometer with the appropriate combination of sensitivity, frequency coverage, and angular resolution. We now briefly summarize propagation phenomena of interest, what they tell us about the medium they traverse, and what is needed to exploit them fully. 

\noindent\textbf{\large Radio Propagation Phenomena}

The solar wind and ISM are comprised of dilute, magnetized, and turbulent plasma. The index of refraction is perturbed from unity and radio waves undergo a variety of ``scattering'' phenomena as a result of their propagation through the medium. A rich phenomenology results that allows a number of plasma properties to be deduced. The table (from [2]) summarizes a number of these phenomena and the plasma property that can be leveraged from observations. One of the most important measurements, particularly for the ISM, is that of angular broadening. As shown in [22] the magnitude of the broadening and its functional form - the ``blurring function'' - can be used to infer the spatial power spectrum of turbulent fluctuations in plasma density. 

\begin{figure}[htp]
\centering
\includegraphics[trim=2.in 1.5in 2.0in 1.5in,clip,width=6.5in]{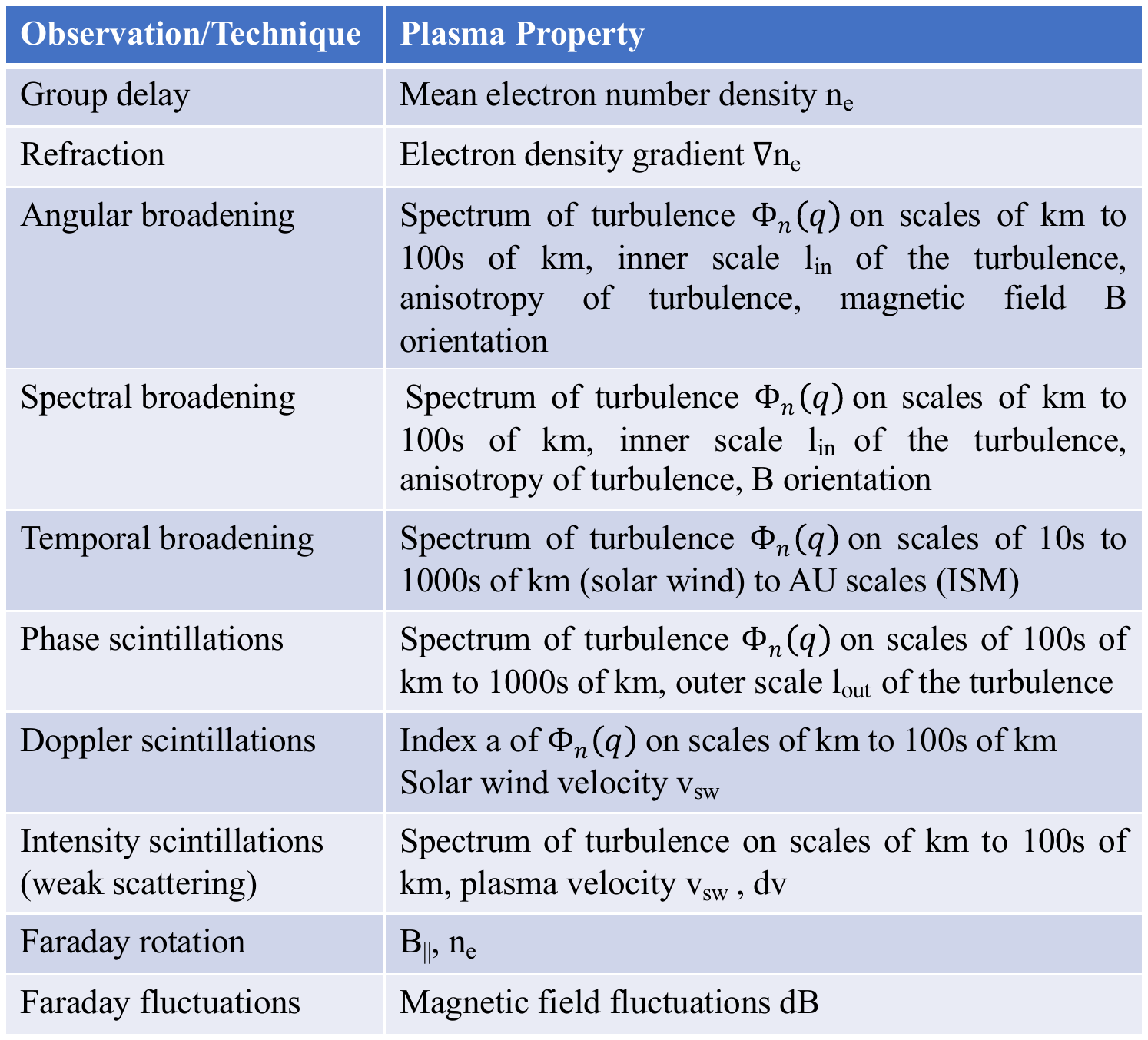}
%\label{fig:Table}
\end{figure}

Observations of radio propagation phenomena requires suitable background sources to probe the foreground medium of interest. In the case of the solar wind, spacecraft beacons have been used, (e.g.,[3,26]), as have radar echos from Venus [13]. More typically, however, cosmic background sources like quasars and pulsars have been exploited. The sensitivity of available instruments (e.g., VLA, VLBA) has been limited and so the number and angular density of such sources that could be used as effective probes of the solar wind and ISM have been low. Moreover, the distribution of interferometric baselines, needed to measure turbulent properties of the medium or its velocity, for example, has also been severely limited. 

\noindent\textbf{\large Requirements for Future Progress}

The use of radio propagation phenomena to leverage detailed information about astrophysical plasmas has been limited primarily by three factors: sensitivity, the distribution of radio interferometric baselines, and time resolution. Sensitivity is needed to observe sufficient numbers of background sources to map the foreground plasma. Antenna baseline coverage is needed to adequately sample the spatial spectrum of the plasma turbulence. High time resolution is needed, in the case of the solar wind, to measure fast and slow solar wind speeds as a function of solar elongation and position angle using intensity scintillations and to exploit pulsar diagnostics. The Jansky Very Large Array (JVLA), re-dedicated in 2012, represents a significant improvement over previous instruments in terms of its sensitivity but its baseline coverage only extends to 35 km.  While the ten-antenna Very Long Baseline Array (VLBA) has baselines of 100s to 1000s of km, its sensitivity is limited. An ideal radio interferometer would have the following attributes for solar wind and ISM studies:

\begin{itemize}
\item High sensitivity, sufficient to allow dense angular sampling of background sources
\item Broad frequency coverage to allow a large range of solar elongations to be sampled
\item Large instantaneous bandwidths for enhanced sensitivity
\item Antenna baselines extending out to ~1000 km with good position angle coverage for angular broadening and scintillation measurements
\item Support of Stokes polarimetry for Faraday rotation measurements
\item Integration times short enough to exploit millisecond pulsars for Faraday rotation and scintillation diagnostics
\end{itemize}

The proposed 244-antenna {\sl next-generation Very Large Array} (ngVLA; [25,28]) will possess all of these attributes and will therefore be ideal for greatly expanded studies of the solar wind and ISM using radio propagation phenomena. The ngVLA will be a fixed array that can monitor solar wind turbulence at all times. In contrast, as a reconfigurable array, the JVLA can only be used on an occasional basis for solar wind studies (roughly 25\% of the time). 

The ngVLA will permit an enormous improvement in this kind of science with respect to the JVLA and VLBA, allowing almost all of the techniques listed in the table to be routinely exploited, the only exceptions being those that require spectrally coherent sources (which will require spacecraft beacons or radar signals). In the case of the solar wind, the ngVLA will measure the spatial spectrum on spatial scales of 10s of meters to ~1000 km, spanning the Kolmogorov-like inertial subrange, the flattened spectrum on scales of km to 100s of km (possibly due to kinetic Alfven waves [6]), and the dissipation range (cf. [9]). Intensity scintillation tomography will allow the fast and slow solar wind velocity field (e.g., [4]) and velocity fluctuations [17] to be mapped in the critical region where the corona transitions to the inner heliosphere, regions that may not be accessible to in situ measurements. Faraday rotation and dispersion measure observations of polarized background sources (e.g., [15]) and pulsars [31] will allow constraints to be placed on the overall density structure and magnetic field of the outer corona and inner heliosphere. We point out that by making global measurements of the solar wind, solar wind turbulence, and transients [18] using radio propagation diagnostics, the ngVLA would be be highly complementary to {\sl in situ} measurements made by next-generation missions such as {\sl Helioswarm}. 

For the interstellar medium, one of the most exciting capabilities of the ngVLA will be the ability to measure, with the same instrument and using the same technique (angular broadening) the interstellar turbulence spectrum over a wide range of spatial wavenumbers, including the spectrum in the dissipation range. We note the intriguing flattening in the Voyager 1 measurements of the ISM turbulence spectrum on kinetic scales [22]. Intensity scintillations in the strong scattering regime, both diffractive and refractive, will also play an important role [1]. The increased sensitivity of the ngVLA relative to the JVLA will render it an ``angular broadening machine'' that can produce measurements of the angular broadening function for a large number of sources. The large range of baseline lengths of the ngVLA will allow much improved measurements of the turbulent density spectrum, as well as detections of a wider angle of scattering strength than is presently possible. This is crucial to an understanding of turbulence, because it indicates the processes responsible for converting turbulent energy to heat. A summary of results for the interstellar medium is given in [14] based on observations made with the VLA, JVLA, and the VLBA. 

Similar to the solar wind, pulsars will also be used by the ngVLA to probe the structure and turbulent properties of the WIM using scintillations and the secondary spectrum (e.g., [29]). The North American Nanohertz Observatory for Gravitational Waves (NANOGrav) has observed dozens of pulsars in recent years, accruing many measurements of the dispersion measure [16]. Careful analyses of DM variations also yield insight into the ISM [19] and the solar wind [24]. Finally, as studies of the so-called fast radio bursts ([23]; see also recent results from the CHIME/FRB Collaboration [7]) mature, they too may be exploited - to study the turbulent properties of the intergalactic medium outside the Milky Way [12].

\end{document}